# Dynamic scaling approach to study time series fluctuations

Alexander S. Balankin

Grupo "Mecánica Fractal", Instituto Politécnico Nacional, México D.F., México 07738

We propose a new approach for properly analyzing stochastic time series by mapping the dynamics of time series fluctuations onto a suitable nonequilibrium surface-growth problem. In this framework, the fluctuation sampling time interval plays the role of time variable, whereas the physical time is treated as the analog of spatial variable. In this way we found that the fluctuations of many real-world time series satisfy the analog of the Family-Viscek dynamic scaling ansatz. This finding permits to use the powerful tools of kinetic roughening theory to classify, model, and forecast the fluctuations of real-world time series.

The internal dynamics of many complex real-world systems is often studied through analysis of the fluctuations of the output time series [1,2]. Statistically, the fluctuations of time series $z(t)$ are commonly characterized by their log-returns $v(t,\tau) = \ln[z(t)/z(t+\tau)]$ for a fixed sampling time interval $\tau$ [3]. A large number of studies have been carried out to analyze the time series of dynamical variables from physical, biological and financial systems (e.g., [1-4,5]). Since the resulting observable variable associated with fluctuations at each moment is the product of log-return magnitude and sign, recent investigations have focused on the study of correlations in the absolute value and sign time series [6]. It was found that the absolute values of log-returns, $V(t,\tau) = |v|$, of many real-world time series exhibit long-range power-law correlations [3-6], indicating that the system does not immediately respond to an amount of information flowing in it, but reacts to it gradually over a period of time [7]. Accordingly, the analysis of the scaling properties of the fluctuations has been shown to give important information regarding the underlying processes responsible for the observed macroscopic behavior of real-world system [1-7].

The long-term memory in the time series fluctuations is commonly analyzed through a study of their structure function, defined as $\sigma(\tau,\delta t) = \left\langle \overline{[V(t+\delta t,\tau) - V(t,\tau)]^2} \right\rangle^{1/2}$, where the overbar denotes average over all $t$ in time series of length $T-\tau$ ($T$ is the length of original time series $z(t)$) and the brackets denote average over different realizations of the time window of size $\delta t$ [3]. Alternatively, one can use power-spectrum of log-return time series, defined as $S(q,\tau) = \langle \hat{V}(q,\tau)\hat{V}(-q,\tau) \rangle$, where $\hat{V}(q,\tau) = (T-\tau)^{-1/2} \sum_t [V(t,\tau) - \overline{V}] \exp(iqt)$ is the Fourier transform of $V(t)$ [3-5]. It was found that the structure function and the power spectrum of the absolute log-returns commonly exhibit the power law behavior

$$\sigma \propto (\delta t)^\zeta \text{ and } S \propto q^{-\theta}, \qquad (1)$$

where $\theta = 2\zeta + 1$, even when the time series $z(t)$ looks erratic and its log-returns $v(t,\tau)$ are apparently random (see [3-5]). The scaling exponent $\zeta$, also called the Hurst exponent, characterizes the strength of long-range correlations in the fluctuation behavior [8]. It has been shown that the knowledge of $\zeta$ is very helpful also for practical purposes [3-7]. However, the scaling behavior (1) characterizes the correlations in a log-return time series treated as an analog of rough interface in (1+1) dimensions (see Refs. [8,9,10]). It's worth nothing that this model gives very limited information about the studied system that does not permit to determine the universality class of the system, even within the framework of an equilibrium-type theory [11]. Additional information can be obtained from the studies of the probability density function of log-return record [12] and/or the diffusion entropy analysis [13].

A better physical understanding of fluctuation dynamics requires a proper description for correlation properties of local variables on different time sampling intervals [5]. In this work, we propose a novel approach to properly understand the dynamics of time series fluctuations by studying the behavior of absolute log-returns for different sampling intervals. Specifically, we suggest to treat the absolute log-return time series as a moving interface $V(t,\tau)$, where the sampling time interval, $\tau$, plays the role of time variable, whereas the physical time, $t$, plays the role of spatial variable.

In this framework, we performed the dynamic analysis of fluctuations in the time series associated with systems of different nature. Specifically, we studied time series associated with the physical, economic, and informatic systems (see Table 1) early analyzed within other frameworks. Namely, the stress-strain behavior $F(\varepsilon)$ associated with the paper damage (see Fig. 1) was studied in [14]. It was found that the damage branch of stress-strain curve itself exhibits persistence associated with the long-range correlations in the fiber network. Radar backscattered signals from different types of soils were studied in [15], where it was shown that the backscattered signals possess a long memory with the Hurst exponent related to the mass fractal dimension of a soil, $D$, as $\zeta = 3 - D$. The records of the daily crude oil price (see Fig 2) and the daily Mexican stock market index early were studied in Refs. [5, 16]. It was found that both time series behave randomly with exponentially decaying auto-correlation function, whereas their fluctuations exhibit long-range power-law correlations. The information flow time series were studied in a large number of works, e.g. [17]. In this work, the data of information flow were taken in WWW IPN server from Ref. [18], where it was found that the fluctuations of information flow in WWW IPN server exhibit scaling behavior (1) with the Hurst exponent $\zeta = 0.5$.

To study the dynamics of time series fluctuations, in this work we analyzed the time series of absolute log-returns of original time series. The lengths ($T$) and the sampling rates ($\Delta t$) of the original time series and the ranges ($\tau_m \leq \tau \leq \tau_M$) and the rates ($\delta \tau$) of the sampling intervals of the absolute log-returns ($\tau$) are reported in the footnotes of Table 1 and in the figure captions. Notice that the length of the log-return time series is $T_{l-r} = T - \tau_M$ (see Fig. 1).

First of all we noted the visual similarity between the log-return time series with different sampling intervals (see Fig. 1 c and 2 b-d). Quantitatively, the self-affinity of fluctuation time series is characterized by the scaling behavior (1), as it is illustrated the graphs in Figs. 3, 4 for stress fluctuations shown in Fig 1 c and in Figs. 5, 6 for the log-returns of crude oil price records shown in Figs 2 b-d. Notice that the structure functions display a power-law behavior (1) with $\zeta(\tau) = const$ within a wide but bounded intervals of $\delta t$ (see Figs. 3 a and 5 a) with the upper cut-off $\delta t_C$ increasing with $\tau$ as

$$\delta t_C \propto \tau^{1/z}, \qquad (2)$$

where $z$ is the dynamic exponent (see Fig. 5 b). Furthermore, we found that the structure function and power spectrum of the absolute log-return also scale with the sampling interval $\tau$ as

$$. \sigma \propto \tau^\beta \text{ and } S \propto \tau^{\theta/z}, \qquad (3)$$

where $\beta$ is the fluctuation growth exponent (see Figs. 3 b, d and 5 c).

The values of scaling exponents for the absolute log-returns of all time series studied in this work are given in Table 1. Notice that in all cases $\theta = 2\zeta + 1$ (see Table 1), as it is expected for self-affine time series (see [8]). Moreover, we found that the dynamic exponent satisfies the scaling relation

$$z = \zeta / \beta, \qquad (4)$$

as it follows from data reported in Table 1.

Our findings indicate the existence of dynamic scaling behavior analogous to the Family-Viscek dynamic scaling ansatz for kinetic roughening of moving interface (see Ref. [10]). Indeed, the scaling relations (1) – (4) imply that the structure function and power spectrum of time series fluctuations exhibit the dynamic scaling behavior

$$\sigma(\tau, \delta t) \propto \tau^{\beta} f_l(\delta t / \tau^{1/z}) \qquad (5)$$

and

$$S \propto q^{-\theta} f_s(q \tau^{1/z}), \qquad (6)$$

respectively, where the scaling functions behave as

$$f_l \propto \begin{cases} y^{\zeta}, & \text{if } y < 1 \\ const, & \text{if } y >> 1 \end{cases} \qquad (7)$$

and

$$f_s \propto \begin{cases} y^\theta, & \text{if } y < 1 \\ const, & \text{if } y \gg 1 \end{cases}, \qquad (8)$$

as it is shown in Figs. 4 a, b, 5 c, 6 b, and 7, where the data collapses are achieved with scaling exponents (see Table 1) obtained with the scaling relations (1)-(3).

It should be pointed out that in the graphs Figs. 4 a, b each point represent the average over 30 experiments, while the data collapses for crude oil price fluctuations shown in Figs 5 c and 6 b corresponds to the single price time series shown in Fig. 2 a. The graph (2) in Fig. 7 also corresponds to data collapse for single time series of the Mexican stock market Index, whereas each point of the graphs (1) and (3) in Fig. 7 correspond to 25 radar backscattered signals and 30 time series of information in WWW IPN server.

Dynamic scaling (5)-(8) permits to treat the fluctuation dynamics as a kinetic roughening of moving interface (visualized in Figs. 1 c and 2 b-d). Therefore, we can use the powerful tools of kinetic roughening theory (see [8-10]) to characterize and model the fluctuations of real-world time series. Specifically, we can try to determine the universality class for fluctuation dynamics associated with the set of two independent scaling exponents, e.g. $\zeta$ and $z$, analogous to the universality class for kinetic roughening phenomena (see [10, 11]).

The dynamical universality classes are determined by the dimensionality, the conservation laws, the symmetry of the order parameter, the range of the interactions, and the coupling of the order parameter to conserved quantities [11]. So, the knowledge of universality class allows us to understand the fundamental processes ruling the system dynamics. Hence, we can construct the kinetic equation governed the fluctuation dynamics. Besides, the representation of the absolute log-return time series as a moving interface permits to use the kinetic equations from the theory of kinetic roughening for the modeling of fluctuation dynamics which belong to the same universality class.

In this way, the time-series fluctuation dynamics is expected to be described by the Langevin-type equation

$$\frac{\partial V}{\partial \tau} = \Phi\left(\frac{\partial V}{\partial t}\right) + \eta(\tau,t) + F, \qquad (9)$$

where $F$ is the external force, while the noise term $\eta(\tau,t)$ and the actual form of the function $\Phi$ are determined by a particular model obeying the corresponding class of universality (see [10]).

In this way, the values of scaling exponents reported in the Table 1 suggest that the fluctuations of radar backscattered signals belong to the Edwards-Wilkinson (EW) universality class of kinetic roughening (see [10]), while the fluctuations of information flow in WWW server belong to Kardar-Parisi-Zhang (KPZ) universality class in 1+1 dimensions (see [10]). So, the fluctuations of information flow in WWW IPN server can be modeled by the seminal KPZ-equation, *i.e.*, Eq. (9) with

$$\Phi = a\left(\frac{\partial^2 V}{\partial t^2}\right) + b\left(\frac{\partial V}{\partial t}\right)^2 \qquad (10)$$

and a Gaussian noise $\eta(\tau,t)$ with zero mean and variance $\langle \eta(\tau,t)\eta(\tau',t')\rangle = \delta(\tau-\tau')\delta(t-t')$, where $a$ and $b$ are the fitting constants and $\delta(.)$ is the delta-function [10]. In the case of EW universality, as it is found for fluctuations of radar backscattering signals, $b = 0$ [10]. The fluctuation dynamics in other systems studied in this work also can be modeled with the help of kinetic roughening theory. Namely, we can use the models of kinetic roughening of interfaces moving in a medium with long-range correlations in the spatio-temporal disorder (see [10,19]). Moreover, we can also model the coupled dynamics of fluctuations in time series associated with coupled systems, using the models of kinetic roughening of coupled interfaces moved in disordered media [20].

Furthermore, making use an appropriate assumptions about the log-return sign dynamics (see [7]), the kinetic roughening models (9) can be used for forecasting of time series of dynamical variables for systems of different nature. Indeed, starting from seminal work of Black & Scholes [21], the most economic time series forecasting model are based on the model representation for future volatility (absolute log-returns) as an appropriate stochastic process, e.g. (fractional) Brownian motion (see, for example, refs. [22] and references therein). Our findings permit to use an appropriate Langevin-type equation (9) to simulate the implied future volatility, when the past realized volatility is known.

In general, the suggested approach may be useful for characterization, modeling, and forecast of stochastic time series arising from diverse science areas, such as statistical physics, econophysics, medicine, and material science.

The author would like to thank O. Susarrey and M. Patiño for technical help and M. A. Rodríguez, and R. Cuerno for useful discussions. This work has been supported by CONACyT of the Mexican Government under Project No. 44722.

Table 1 Characteristic scaling exponents of absolute log-return time series and the corresponding universality classes of fluctuation dynamics (EW – Edwards-Wilkinson and KPZ – Kardar-Parisi-Zhang universality classes in 1+1 dimensions; see [10]).

| Time series | $\zeta$ from eq. (1) | $\beta$ from eq. (3) | $\theta$ from eq. (1) | $z$ from eq. (2) | $z = \zeta/\beta$ eq. (4) | Class |
|---|---|---|---|---|---|---|
| Stress fluctuations[1] (see Fig. 1) | 0.39±0.02 | 0.55±0.05 | 1.8±0.2 | 0.65±0.1 | 0.7±0.07 | ? |
| Radar backscattered signals [2] | 0.50±0.03 | 0.25±0.03 | 2±0.2 | 1.8±0.4 | 2±0.06 | EW |
| Crude oil price [3] (see Fig. 2) | 0.36±0.02 | 0.50±0.03 | 1.7±0.1 | 0.7±0.1 | 0.72±0.05 | ? |
| Mexican stock market Index [4] | 0.49±0.02 | 0.63±0.03 | 2.1±0.02 | 0.7±0.1 | 3/4±0.05 | ? |
| Information flow in WWW IPN server [5] | 0.50±0.04 | 0.34±0.02 | 1.9±0.2 | 1.6±0.4 | 3/2±0.6 | KPZ |

[1] The force series with the sampling rate $\Delta t = 0.2$ sec. have the length $T = 900 \pm 100$ second (see [14]); in this work we studied the absolute log-returns with the sampling interval $\tau$ varied in the range from 10 to 100 sec. with the rate of 10 sec. Data reported for stress fluctuations are the average over results obtained for 30 stress-strain curves from [14]; accordingly the error range is determined by the standard variation of data for different curves.

[2] Radar backscattered signals from different types of soils were studied in [16]. In this work we study the fluctuations of these signals of the length $T = 50$ nanoseconds with the sampling rate of 0.01 nanoseconds (see [16]). Reported data are the average over results obtained for 25 signals.

[3] Time record of West Texas Intermediate crude oil spot price in the 2003 constant dollars per barrel was taken from the Bloomberg database [23]. The length of record is $T = 4879$ business days (from 01.02.1986 to 04.26.2005).

[4)] The record of Mexican stock market daily index for period of 4438 business days (from 01.04.1988 to 09.23.2005) was taken from [24]).

[5)] The information flow in the WWW IPN server records of length $T = 7000$ second with the sampling rate $\Delta t = 1$ sec. were taken from [18]. Reported data are the average over results obtained for 30 records.

**Figure captions**

**Figure 1**. (a) Damage image of toilet paper sheet of width $L = 10$ cm under the uniaxial tension stress with the constant displacement rate $\dot{u} = 0.01$ mm/sec.; (b) the corresponding times series of stress $F(t_f)$ of length $T = 650$ sec. with the sampling rate of 0.2 sec., related to the stress-strain curve $F(\varepsilon) = F(\varepsilon = \dot{u}t/L)$ shown in the insert (see Ref. [16]); and (c) the stress fluctuations $V(t_V, \tau)$ for different sampling time intervals in the range $10 \leq \tau \leq 100$ sec. (notice that $t_V = t_f - \max \tau = t_f - 100$ sec, where $t_f = 0$ corresponds to the local maximum of the decreasing branch of stress-strain curve $F(\varepsilon) = F(\dot{u}t/L)$ (see insert in the panel b), so the lengths of all series $V(t_V, \tau)$ is 750 sec., with the sampling rate of 0.2 sec.).

**Figure 2**. (a) Time series of crude oil price in the 1983 constant US dollars per barrel (from 01.02.1986 to 04.26.2005; source: Bloomberg database [23]) and (b-d) its absolute log-returns for the period of 4096 business days (from 10.16.1986 to 12.20.2002) with the sampling time intervals $\tau = $ 25 (b), 50 (c), and 100 (d) business days.

**Figure 3.** Scaling behavior of: (a, b) structure function and (c, d) power law spectrum of stress fluctuations (see Fig. 1 c). a) $\sigma(\tau, \delta t)$ (arbitrary units) versus $\delta t$ ($10 \leq \delta t \leq 350$ sec.) for $\tau = $ 5 (1), 40 (2), and 160 second (3); straight lines – data fitting by the first Eq. (1); notice that curves are shifted along y-axis for clarity; b) $\sigma(\tau, \delta t)$ (arbitrary units) versus $\tau$ (second) for $\delta t = $ 100 (1) and 200 second (2); straight lines – data fitting by the first Eq. (3); c) $S$ (in arbitrary units) versus $q$ (1/sec.) for $\tau = 20$ sec. (graph obtained with the help of BENOIT 1.3 software [25]); straight line – data fitting by the second Eq. (1); and d) $S$ (arbitrary units) versus $\tau$ (second) for $q = 1/\delta t = 0.01$ sec.$^{-1}$; straight line – data fitting by the second Eq. (3).

**Figure 4**. Data collapses (in arbitrary units) for: a) structure function and b) power spectrum ($S^* = 10^4 S$) of the absolute log-returns of stress fluctuations (see Fig. 1 c) with the sampling intervals $\tau =$ 160 (1), 100 (2), 75 (3), 50 (4), 25 (5), and 10 second (6) for the time windows varying in the range $15 \leq 1/q = \delta t \leq 800$ seconds. Straight lines – data fitting with Eqs. (5,7) and (6,8), respectively; notice that in both graphs the gray tone symbols denote the data excluded from the power-law fitting.

**Figure 5**. Scaling behavior of: the structure function of absolute log-returns of crude oil price: a) $\sigma(\tau, \delta t)$ (arbitrary units) versus $\delta t$ (business days) for $\tau =$ 13 (1), 25 (2), 50 (3), 100 (4), 150 (5) and 198 (6) business days (straight lines data fitting with Eq. (1); curves are shifted along y-axis for clarity); b) $\delta t_C$ (business days) versus $\tau$ (business days), straight lines data fitting with Eq. (2); c) $\sigma(\tau, \delta t)$ (arbitrary units) versus $\tau$ (business days) for $\delta t =$ 300 (1) and $T =$ 4096 (2) business days (straight lines data fitting with Eq. (2); full symbols are excluded from poer fitting); and c) data collapse (in arbitrary units) for series with sampling intervals $\tau =$ 10 (1, 2), 25 (3, 4), 150 (5, 6), and 198 (7, 8) business days for the time windows varying in the range $3 \leq \delta t \leq 300$ business days, straight lines – data fitting with Eqs. (5, 7).

**Figure 6**. Scaling behavior of: the power spectrum of absolute log-returns of crude oil price: a) $S$ (arbitrary units) versus $q$ (1/business day) for sampling intervals $\tau$ = 50 (1) and 198 (2) business days (graphs obtained with the help of BENOIT 1.3 software [25]); straight lines – data fitting with the second Eq. (1); notice that the graphs are shifted for clarity and the gray tone symbols denote the data excluded from the power-law fitting. b) Data collapse (in arbitrary units) for the power spectrum of time series with the sampling intervals $\tau$ = 50 (1, 2), 70 (3, 4), 100 (5, 6), and 198 (7, 8) business days for the time windows varying in the range $10 \leq 1/q = \delta t \leq 2000$ business days; straight lines denotes the data collapse according to Eqs. (6, 8); the slope of dashed line is 1.72. Insert shows the power law fitting of data of series (1, 3, 5, 7) denoted by the same symbol (circle).

**Figure 7**. Data collapse (in arbitrary units) for the structure function of absolute log-returns of: a) radar backscattered signals with the sampling intervals $\tau$ = 0.1 (1, 2), 1 (3, 4), and 5 ns (5, 6) ns for the time windows varying in the range $0.1 \leq \delta t \leq 4$ ns; b) Mexican stock market index with the time intervals $\tau$ = 10 (1, 2), 43 (3, 4), 100 (5), and 180 business days (6) for the time windows varying in the range $3 \leq \delta t \leq 160$ business days; and c) information flow in WWW IPN server with the sampling intervals $\tau$ = 10 (1, 2), 100 (3, 4), and 360 second (5) for the time windows varying in the range $10 \leq \delta t \leq 300$ second. Straight lines – data fitting with Eqs. (5, 7).

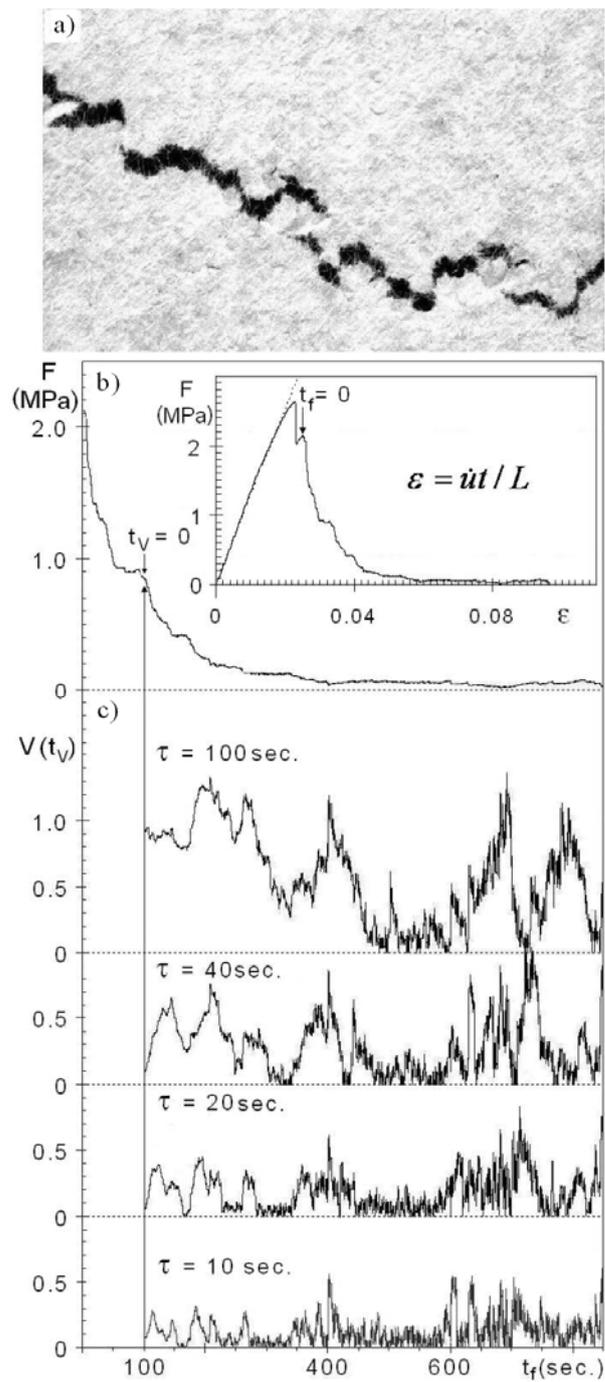

Figure 1.

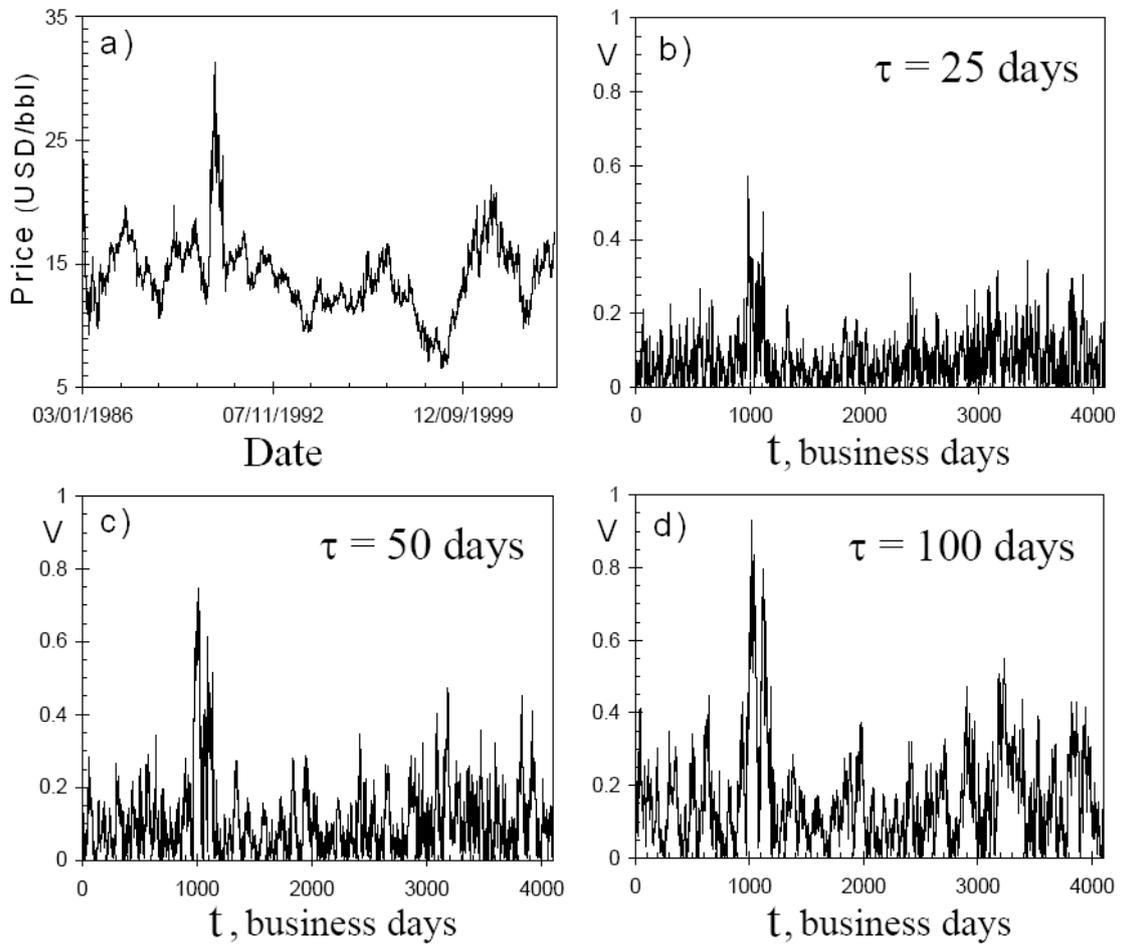

Figure 2.

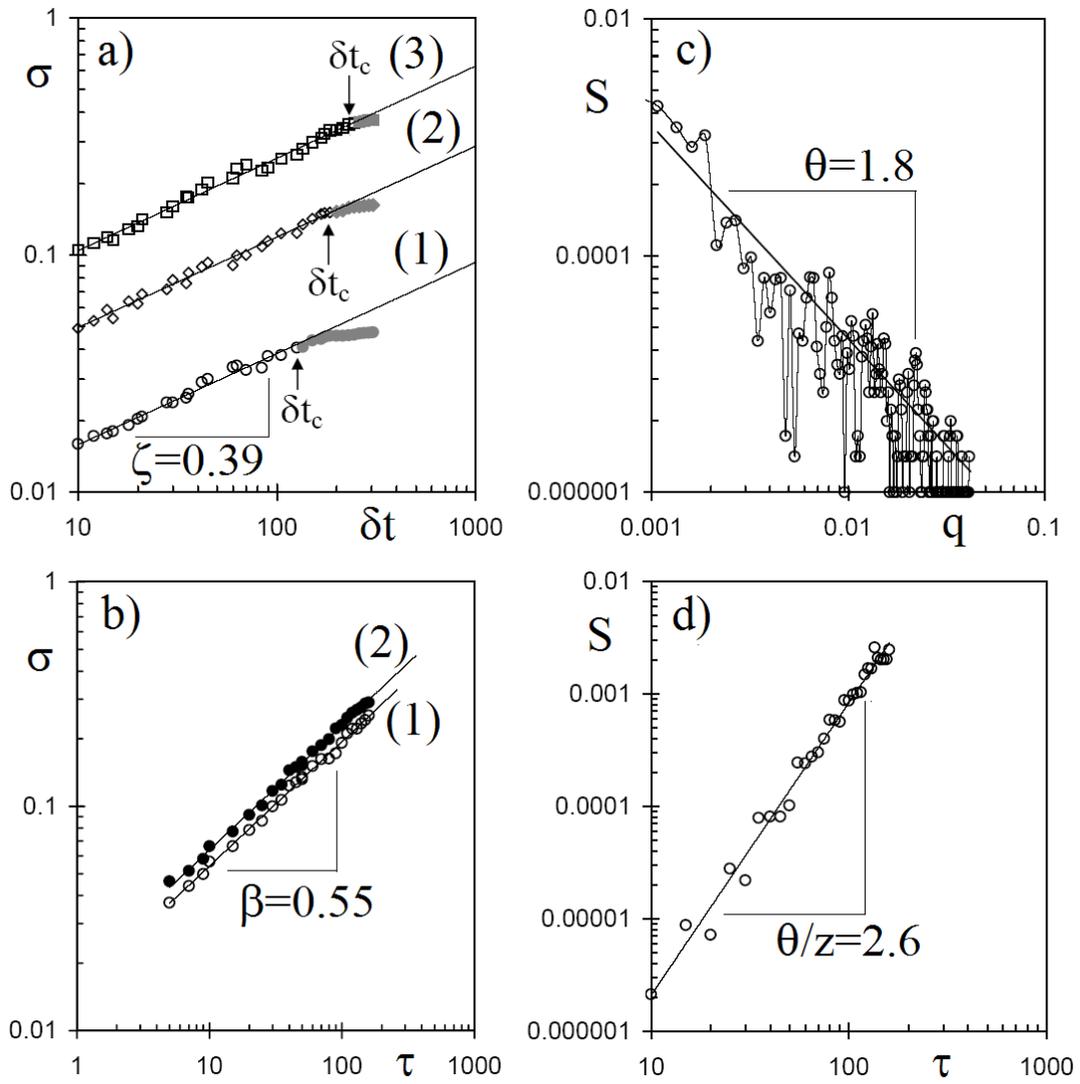

Figure 3.

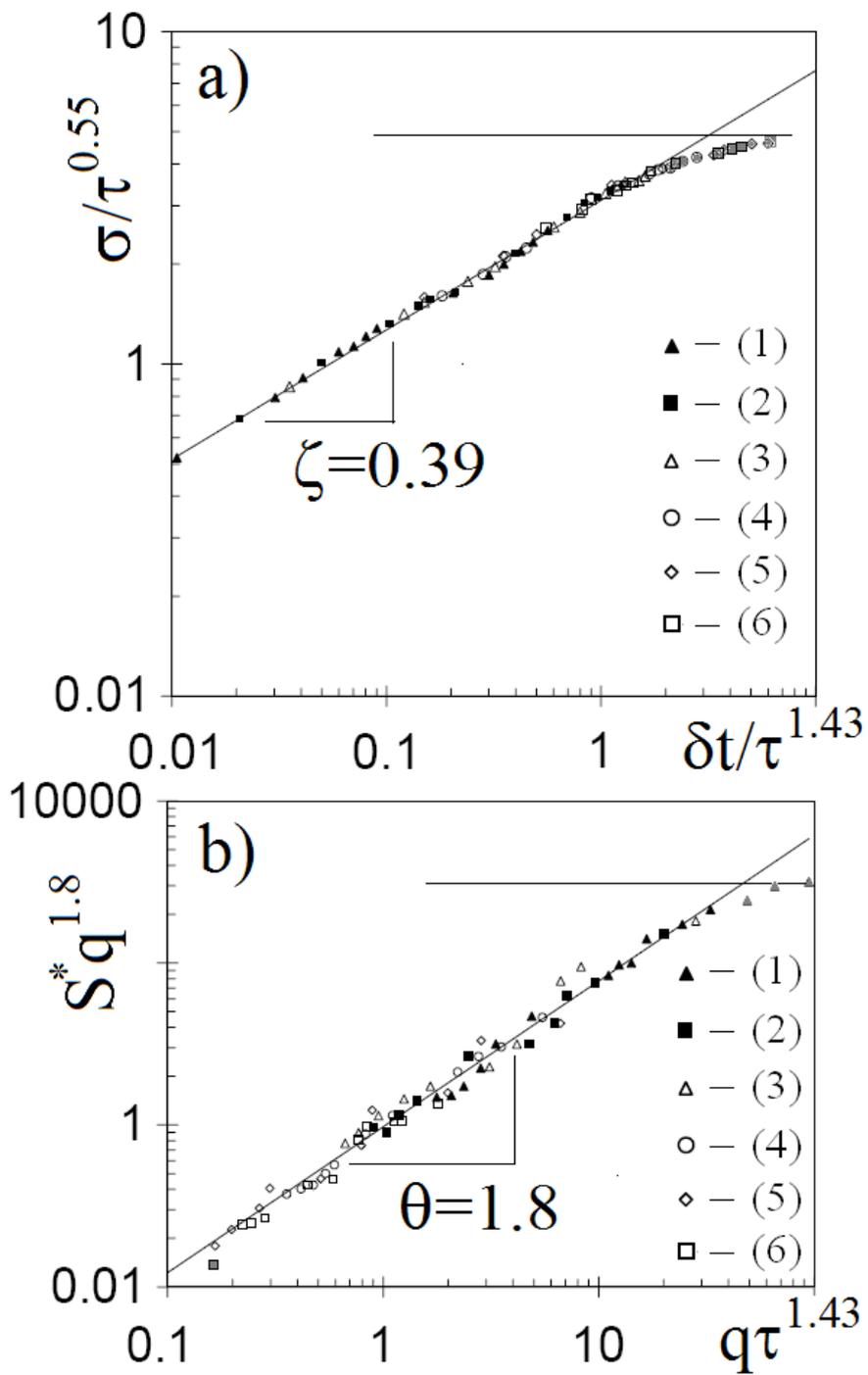

Figure 4.

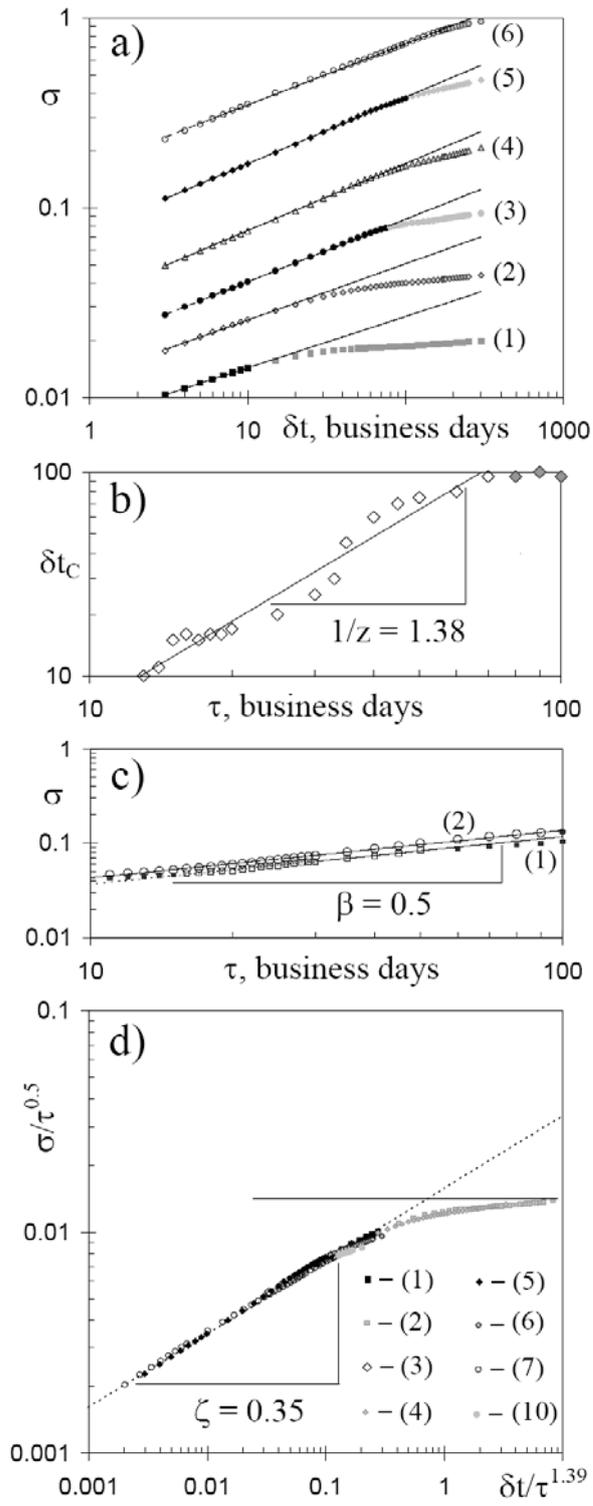

Figure 5.

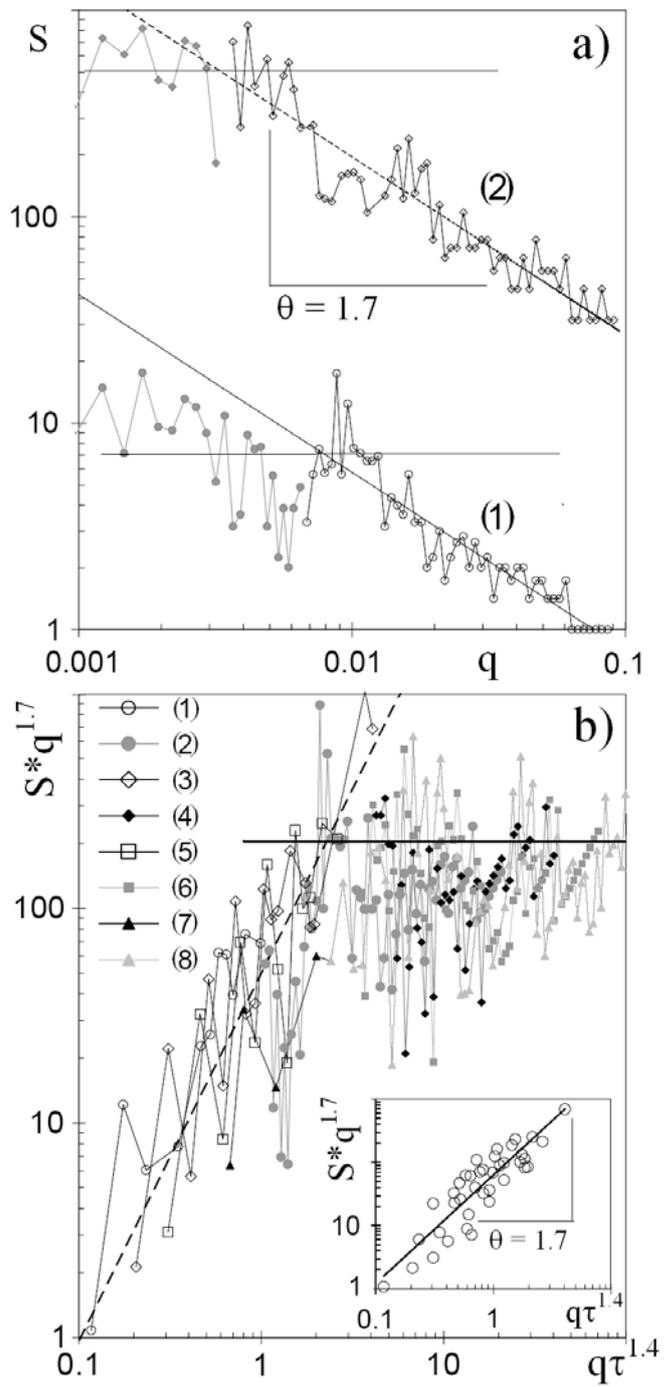

Figure 6.

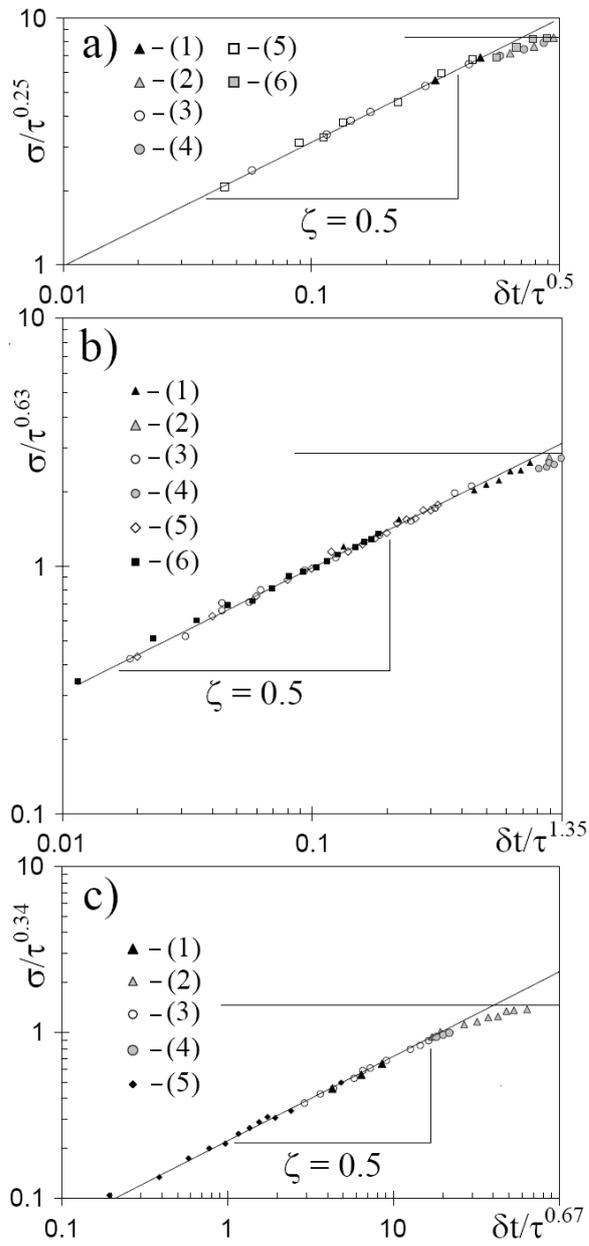

Figure 7.